\documentstyle[pre,aps,epsfig,rotate,preprint]{revtex} 
\begin{document}                                     
\title{Modeling and Simulation of Multi-Lane Traffic Flow}
\author{Dirk Helbing and Andreas Greiner}
\address{II. Institute of Theoretical Physics, University of
Stuttgart, 70550 Stuttgart, Germany}
\maketitle
\vfill                                                    
\begin{abstract}
A most important aspect in the field of traffic modeling is the simulation
of bottleneck situations. For their realistic description a 
macroscopic multi-lane model for uni-directional freeways including
acceleration, deceleration, velocity fluctuations, overtaking and
lane-changing maneuvers is systematically deduced from a gas-kinetic
(Boltzmann-like) approach. The resulting equations contain corrections
with respect to previous models. For efficient computer simulations,
a reduced model delineating the coarse-grained temporal behavior
is derived and applied to bottleneck situations.\\[3mm]
\end{abstract}                                                
\vfill
\clearpage

\section{Introduction}
Apart from microscopic traffic models, 
in the last decades a number of interrelated macroscopic traffic models have
been proposed 
\cite{LiWhi55,Ri56,Pay71,Pay79b,Kue84,KueRoe91,KeKo93,KeKo94,Hel95,Hel96}.
The motivations for developing these were
\begin{itemize}
\item to describe and understand the instabilities of traffic flow
\cite{Kue84,KueRoe91,KeKo93,KeKo94,Hel96,KueBe93,Hel95c}, 
\item to optimize traffic flow by means of on-line speed-control
systems \cite{Kue87,KueKr92,Hel95d},
\item to make short-term forecasts of traffic volumes for re-routing
measures \cite{HiReWe93,Hil95,Hil96},
\item to calculate the average travel times, fuel consumption, and
vehicle emissions in dependence of traffic volume 
\cite{Pay79b,Hel97},
\item to predict the effects of additional roads or lanes 
\cite{Hel97,Bra68,Bas92,Kno93}. 
\end{itemize} 
Most of these models are restricted to uni-directional freeway traffic and 
treat the different lanes of a road in an overall manner, i.e. 
like one lane with higher capacity and possibilities for overtaking.
However, this kind of simplification is clearly not applicable 
if there is a disequilibrium between neighboring lanes. 
Therefore, some researchers carried out empirical
investigations of the observed density oscillations between neighboring lanes
or proposed models for their mutual influences 
\cite{GaHeWe62,MuPi71,MuHs71,Ror76,MaNa81,MiBe84}.
\par
However, these are phenomenological models which
treat inter-lane interactions in a rather heuristic
way. Moreover, most of them base on the simple traffic flow model of 
Lighthill and Whitham which assumes average velocity on each lane
to be in equilibrium with density. This assumption is not very well justified,
especially for unstable traffic which is characterized by evolving
stop-and-go waves \cite{Kue84,KueRoe91,KeKo93,KeKo94,Hel96,KueBe93,Hel95c}.
It is also questionable for lane mergings or on-ramp traffic where
frequently a disequilibrium occurs \cite{Kue84,KeKo94}.
However, instabilities or disequilibria may decrease the freeway capacity
considerably. 
\par
Another approach including a phenomenological velocity equation
has been proposed by Michalopoulos et al.
\cite{MiBe84}. It bases on Payne's model \cite{Pay71,Pay79b} which 
has been severely criticized for several reasons
\cite{Kue84,Hel96,HaHu79,Pay79,Pap83,Cre85,Smu87,RaLi87}.
Therefore, we will derive a consistent macroscopic multi-lane model 
from a {\em gas-kinetic} level of description.
This is related to Paveri-Fontana's approach (cf. Sec. II), but explicitly 
takes into account overtaking and lane-changing maneuvers. 
The corresponding Boltzmann-like model allows to deduce
macroscopic traffic equations not only for the vehicle densities 
on the different lanes, but also 
for the associated average velocities (cf. Sec. III). 
Due to different legal regulations, the traffic dynamics on American
freeways is different from that on European ones (which will be called
{\em ``autobahns''} in accordance with K\"uhne et al. \cite{Kue87,KueRoe91}).
\par
For efficient computer simulations of large parts of a freeway system
it is desirable to have a somewhat simpler model. Therefore, in Section
IV we will eliminate the velocity equations and derive a reduced 
multi-lane model for the traffic dynamics on a slow time
scale. By means
of computational results it is demonstrated that even the difficult bottleneck
situations can be successfully simulated with this model.
\par
A summary and outlook is presented in Sec. V. However, before 
the {\em macroscopic} multi-lane model sketched in Ref.~\cite{Gra} is 
founded, derived, simplified, and simulated, the
alternative {\em microscopic} approaches shall be mentioned. 
One class of microsimulation models
bases on cellular automata, like the ones by Rickert et al. \cite{Rick}
and Nagatani \cite{Nagat}. These update the vehicle dynamics within two
successive steps: either the vehicle motion in one step and lane-changing in
the next step \cite{Rick}, or the left lane in the 
first step and the right lane in the
second one \cite{Nagat}. Bottlenecks were represented by crashed cars with
zero velocity \cite{Nagat}. Noteworthy are also the event-oriented
model by Wiedemann and Benz \cite{Benz} and the social force model
by Helbing and Schwarz \cite{Hel97}.

\section{Boltzmann-like multi-lane theory}

The first Boltzmann-like (gas-kinetic) model was proposed by Prigogine and
co-workers \cite{PrAn60,Pri61,PrHe71}. However, Paveri-Fontana \cite{Pa75} 
has pointed out that this model has some peculiar 
properties. For this reason, Paveri-Fontana proposed an improved model that
overcomes most of the short-comings of 
Prigogine's approach. Nevertheless, his model still treats
the lanes of a multi-lane road in an overall manner. 
Therefore, an extended Paveri-Fontana-like model will now be constructed.
\par
Let us assume that the motion of an individual vehicle $\alpha$ can be
described by several variables like its {\em lane} $i_\alpha(t)$,
its {\em place} $r_\alpha(t)$, its {\em actual velocity}
$v_\alpha(t)$, and its {\em desired velocity} $v_{0\alpha}(t)$
in dependence of time $t$.
The {\em phase-space density}
$\hat{\rho}_i(r,v,v_0,t)$
is then determined by the mean number $\Delta 
n_i(r,v,v_0,t)$ of vehicles on {\em lane} $i$ that are at a
place between $r - \Delta r/2$ and $r + \Delta r/2$, 
driving with a velocity between $v - \Delta v/2$ and $v + \Delta v/2$, and
having a desired velocity between $v_0 - \Delta v_0/2$ and $v_0 + \Delta v_0/2$
at time $t$:
\begin{eqnarray}
 \hat{\rho}_i(r,v,v_0,t) 
 &=& \frac{\Delta n_i(r,v,v_0,t)}{\Delta r \, \Delta v \, \Delta v_0} 
 \nonumber \\
 & & \hspace*{-1.5cm} =
  \frac{1}{\Delta r \, \Delta v \, \Delta v_0} \sum_\alpha 
 \delta_{ii_\alpha(t)} 
 \!\!\!\!\!\int\limits_{r-\Delta r/2}^{r+\Delta r/2}\!\!\!\!\! dr'
 \delta( r' - r_\alpha(t))
 \!\!\!\!\!\int\limits_{v-\Delta v/2}^{v+\Delta v/2}\!\!\!\!\! dv'
 \delta( v' - v_\alpha(t)) 
 \!\!\!\!\!\int\limits_{v_0-\Delta v_0/2}^{v_0+\Delta v_0/2}\!\!\!\!\! dv'_0 \,
  \delta( v'_0 - v_\alpha^0(\!\not t)) \, . 
 \nonumber \\
 & & 
\end{eqnarray}
Here, $\Delta r$, $\Delta v$, and $\Delta v_0$ are small
intervals. $\delta_{ij}$ denotes the Kronecker symbol and
$\delta(x-y)$ Dirac's delta function. The notation ``$\not t$'' 
indicates that a time-dependence only occurs in exceptional cases.
Lane numbers $i$ are counted in increasing order from the
right-most to the left-most lane, but in Great Britain and Australia
the other way round. (For Great Britain
and Australia ``left'' and ``right'' must always be interchanged.) 
\par
Now we utilize the fact that, due to the conservation of the number of
vehicles, the phase-space density $\hat{\rho}_i(r,v,v_0,t)$ on lane $i$
obeys the {\em continuity equation} \cite{Hel97,Pa75,AlBe78}
\begin{eqnarray}
 \frac{\partial \hat{\rho}_i}{\partial t} + 
 \frac{\partial}{\partial r} ( \hat{\rho}_i v )
 + \frac{\partial}{\partial v} ( \hat{\rho}_i f_i^0 )
 &=& \left( \frac{\partial \hat{\rho}_i}{\partial t} \right)_{\rm ad}
 \!\! + \left( \frac{\partial \hat{\rho}_i}{\partial t} \right)_{\rm int}
 \!\! + \left( \frac{\partial \hat{\rho}_i}{\partial t} \right)_{\rm lc}
 \nonumber \\[2mm]
 &+& \hat{\nu}_i^+(r,v,v_0,t) - \hat{\nu}_i^-(r,v,v_0,t) \, .
\label{Boltz1}
\end{eqnarray}
The second and third term describe temporal changes of the phase-space
density $\hat{\rho}_i(r,v,v_0,t)$ due to changes $dr/dt = v$
of place $r$ and due to acceleration $f_i^0$, respectively.
We will assume that the vehicles accelerate to their desired velocity 
$v_0$ with a certain, density-dependent {\em relaxation time} $\tau_i$,
so that we have the {\em acceleration law}
\begin{equation}
 f_i^0(r,v,v_0,t) = \frac{v_0 - v}{\tau_i} \, .
\label{acc}
\end{equation}
The terms on the right-hand side of equation (\ref{Boltz1}) reflect changes
of phase-space density $\hat{\rho}_i(r,v,v_0,t)$ due to 
{\em discontinuous} changes of desired velocity $v_0$, actual velocity $v$, 
or lane $i$. $\nu_i^+(r,v,v_0,t)$ and $\nu_i^-(r,v,v_0,t)$ are the {\em rates
of vehicles entering and leaving the road} at place $r$. They are only
different from zero for merging lanes at entrances and exits respectively.
\par
The term
\begin{equation}
 \left( \frac{\partial \hat{\rho}_i}{\partial t} \right)_{\rm ad} 
 = \frac{\tilde{\rho}_i(r,v,t)}{T_{\rm r}} [ \hat{P}_{0i}(v_0;r,\!\not t)
 - P_{0i}(v_0;r,t) ] \, ,
\label{adj}
\end{equation}
where 
\begin{equation}
 \tilde{\rho}_i(r,v,t) = \int dv_0 \, \hat{\rho}_i(r,v,v_0,t)
\end{equation}
is  a {\em reduced phase-space density} and $T_{\rm r} \approx 1$\,s is
about the {\em reaction time}, describes an adaptation of the {\em actual
distribution of desired velocities} $P_{0i}(v_0;r,t)$ 
to the {\em reasonable distribution of desired velocities}
$\hat{P}_{0i}(v_0;r,\!\not t)$ without any related change of actual velocity $v$. 
\par
For the reasonable distribution of desired velocities we will assume the
functional dependence
\begin{equation}
 \hat{P}_{0i}(v_0;r,\!\not t) = {\textstyle\frac{1}{\sqrt{2\pi
 \hat{\theta}_{0i}}}} \mbox{e}^{-[v_0 - \hat{V}_{0i}]^2/[2 \hat{\theta}_{0i}]}
\end{equation}
which corresponds to a normal distribution and is empirically
well justified \cite{Pam55,MuPi71,Hel97}. The mean value 
$\hat{V}_{0i} = \hat{V}_{0i}(r,\!\not \!t)$ and variance $\hat{\theta}_{0i}
= \hat{\theta}_{0i}(r,\!\not \!t)$ of $\hat{P}_{0i}(v_0;r,\!\not \!t)$ depend
on road conditions and speed limits. 
Since European autobahns usually do not have speed limits (at least
in Germany), $\hat{\theta}_{0i}$ is larger for these than for American 
freeways. In addition, on European autobahns
$\hat{V}_{0i}$ increases with increasing lane number $i$ since
overtaking is only allowed on the left-hand lane. 
\par
Before we specify the Boltzmann-like interaction term $(\partial \hat{\rho}_i
/ \partial t)_{\rm int}$ and the lane-changing term $(\partial \hat{\rho}_i
/\partial t)_{\rm lc}$ we will discuss some preliminaries. For reasons of
simplicity we will only treat vehicle interactions within the {\em same} lane 
as {\em direct pair interactions,} i.e. in a Boltzmann-like manner
\cite{QSoz}. Lane-changing maneuvers of impeded vehicles that want to escape a queue
(i.e. leave and overtake it) may depend on interactions of up to six vehicles
(the envisaged vehicle, the vehicle directly in front of it, and up to
two vehicles on both neighboring lanes which may prevent overtaking
if they are too close). Therefore, we will treat lane-changing maneuvers
in an overall manner by specifying overtaking probabilities and waiting times of 
lane-changing maneuvers (which corresponds to a {\em mean-field approach},
cf. Ref.~\cite{QSoz}).
These probabilities and waiting times dependent on the
vehicle densities and maybe also on other quantities.
\par
For not too large vehicle densities 
the Boltzmann-like interaction term can be written in the form \cite{Hel97}
\begin{mathletters}\label{Boltz2}\begin{eqnarray}
 \left( \frac{\partial \hat{\rho}_i}{\partial t} \right)_{\rm int}
 &=& \sum_{i'} \int dv' \!\!\int\limits_{w < v'} \!\! dw \int dw_0 \,
 W_2(v,i|v',i';w,i') \hat{\rho}_{i'}(r,v',v_0,t) \hat{\rho}_{i'}(r,w,w_0,t)
 \label{Boltz2a} \\
 &-& \sum_{i'} \int dv' \!\!\int\limits_{w < v} \!\! dw \int dw_0 \,
 W_2(v',i'|v,i;w,i) \hat{\rho}_i(r,v,v_0,t) \hat{\rho}_i(r,w,w_0,t) \, .
\label{Boltz2b}
\end{eqnarray}\end{mathletters}
Term (\ref{Boltz2a}) describes an increase of phase-space density 
$\hat{\rho}_i(r,v,v_0,t)$ by interactions of a vehicle with actual velocity
$v'$ and desired velocity $v_0$ on line $i'$ 
with a slower vehicle with actual velocity
$w < v'$ and desired velocity $w_0$ causing the former vehicle to 
change its velocity to $v \ne v'$ or its lane to $i\ne i'$. The frequency
of such interactions is proportional to the phase-space density 
$\hat{\rho}_{i'}(r,w,w_0,t)$ of hindering vehicles and the phase-space density
$\hat{\rho}_{i'}(r,v',v_0,t)$ of vehicles which can be affected by slower
vehicles. Analogously, term (\ref{Boltz2b}) describes 
a decrease of phase-space density $\hat{\rho}_i(r,v,v_0,t)$ 
by interactions of a vehicle with actual velocity
$v$ and desired velocity $v_0$ on line $i$ 
with a slower vehicle with actual velocity
$w < v$ and desired velocity $w_0$ causing the former vehicle to 
change its velocity to $v' \ne v$ or its lane to $i'\ne i$. Since the
interaction is assumed not to influence the desired velocities $v_0$,
$w_0$, the interaction rate $W_2$ is independent of these. However, the
interaction rate $W_2(v',i'|v,i;w,i)$ is proportional to the relative
velocity $|v-w|$ of approaching vehicles. Therefore, we have the following
relation:
\begin{mathletters}\label{intrate}\begin{eqnarray}
 W_2(v',i'|v,i;w,i) &=& p_i^+ |v-w|
 \delta_{i'(i+1)} \delta(v' - v) \label{inta} \\
 &+& p_i^- |v-w| \delta_{i'(i-1)}
 \delta(v' - v) \label{intb} \\
 &+& (1 - p_i) |v - w|
 \delta_{i'i} \delta(v' - w) \, . \label{intc} 
\end{eqnarray}\end{mathletters}
Term (\ref{inta}) describes an {\em undelayed overtaking} on 
lane $i' = i+1$ without any change of velocity ($v' = v$)
by vehicles which would be hindered by slower 
vehicles on lane $i$. $p_i^+$ denotes the corresponding probability of
immediate overtaking. Analogously, term (\ref{intb}) reflects undelayed
overtaking maneuvers on lane $i' = i-1$ with probability $p_i^-$. Term 
(\ref{intc}) with 
\begin{equation}
 p_i = p_i^+ + p_i^-
\end{equation}
delineates situations where a vehicle cannot be immediately
overtaken by a faster vehicle so that the latter must stay on
the same lane ($i' = i$) and decelerate to the velocity $v' = w$ of the
hindering vehicle. 
\par
We come now to the specification of the lane-changing term $(\partial
\hat{\rho}_i/\partial t)_{\rm lc}$. This has the form of a master equation:
\begin{mathletters}\label{spont}\begin{eqnarray}
 \left( \frac{\partial \hat{\rho}_i}{\partial t}\right)_{\rm lc}
 &=& \sum_{i'(\ne i)} W_1(i|i') \hat{\rho}_{i'}(r,v,v_0,t) \label{spona} \\
 &-& \sum_{i'(\ne i)} W_1(i'|i) \hat{\rho}_{i}(r,v,v_0,t) \, . \label{sponb}
\end{eqnarray}\end{mathletters}
Term (\ref{spona}) describes an increase of phase-space density
$\hat{\rho}_{i}(r,v,v_0,t)$ due to changes from lane $i' \ne i$ to lane $i$
by vehicles with actual velocity $v$ and desired velocity $v_0$. 
The frequency of lane-changing
maneuvers is proportional to the phase-space density
$\hat{\rho}_{i'}(r,v,v_0,t)$ of vehicles which may be interested in
lane-changing. Analogously, term (\ref{sponb}) reflects changes from lane $i$
to another lane $i'$ causing a decrease of $\hat{\rho}_{i}(r,v,v_0,t)$. 
For the corresponding rate $W_1(i'|i)$ of lane-changing maneuvers we 
have the relation
\begin{equation}
 W_1(i'|i) = \frac{1}{T_i^+} \delta_{i'(i+1)} + \frac{1}{T_i^-}
 \delta_{i'(i-1)} \, ,
\label{w1}
\end{equation}
since vehicles can only change to the neighboring lanes $i' = i\pm 1$.
$T_i^+$ [$T_i^-$] denotes the {\em waiting times} for {\em delayed} overtaking or
spontaneous lane-changing maneuvers on the left-hand [right-hand] lane.
\par
Due to different legal regulations, the explicit form of the
overtaking probabilities $p_i^\pm$ and the waiting times $T_i^\pm$
in dependence of the vehicle densities is different in American countries
compared to European ones. A more detailled discussion of this aspect
is presented in Ref.~\cite{Hel97}.

\section{Derivation of macroscopic traffic equations}

The gas-kinetic traffic equations are not very suitable for computer
simulations since they contain too many variables. Moreover, the phase-space
densities are very small quantities and, therefore, subject to considerable
fluctuations so that a comparison with empirical data is difficult.
However, the special value of gas-kinetic traffic equations is that they
allow a systematic derivation of dynamic 
equations for the macroscopic (collective) quantities 
one is mainly interested in. 

\subsection{Definition of Variables}

The most relevant macroscopic quantities are the {\em vehicle densities}
\begin{equation}
 \rho_i(r,t) = \int dv \int dv_0 \, \hat{\rho}_i(r,v,v_0,t)
\end{equation}
and the {\em average velocities}
\begin{equation}
 V_i(r,t) \equiv \langle v \rangle_i = \int dv \, v P_i(v;r,t) 
\end{equation}
on lanes $i$. Here, we have applied the notation
\begin{equation}
 F_i(r,t) \equiv \langle f(v,v_0) \rangle_i = \int dv \int dv_0 \,
 f(v,v_0) \frac{\hat{\rho}_i(r,v,v_0,t)}{\rho_i(r,t)} 
\end{equation}
and introduced the {\em distribution of actual velocities}
\begin{equation}
 P_i(v;r,t) = \int dv_0 \, \frac{\hat{\rho}_i(r,v,v_0,t)}{\rho_i(r,t)}
 = \frac{\tilde{\rho}_i(r,v,t)}{\rho_i(r,t)}
\end{equation}
on lane $i$.
Analogous quantities can be defined for vehicles entering and leaving
the road at entrances and exits, respectively.
\begin{equation}
 \nu_i^\pm(r,t) = \int dv \int dv_0 \, \nu_i^\pm(r,v,v_0,t)
\end{equation}
are the {\em rates of entering and leaving vehicles}, and
\begin{equation}
 V_i^\pm(r,t) \equiv \langle v \rangle_i^\pm = \int dv \, v
 P_i^\pm(v;r,t)
\end{equation}
their {\em average velocities}, where
\begin{equation}
 P_i^\pm(v;r,t) = \int dv_0 \, \frac{\hat{\nu}_i^{\pm}(r,v,v_0,t)}
 {\nu_i^\pm(r,t)}
\end{equation}
are the {\em velocity distributions} of entering and leaving 
vehicles, respectively.
In addition, we will need 
the {\em velocity variance}
\begin{equation}
 \theta_i(r,t) \equiv \langle [v - V_i(r,t)]^2 \rangle_i
 = \int dv \, [v - V_i(r,t)]^2 P_i(v;r,t) = \langle v^2 \rangle_i
 - (\langle v \rangle_i)^2
\end{equation}
and the {\em average desired velocity}
\begin{equation}
 V_{0i}(r,t) = \int dv \int dv_0 \, v_0 \frac{\hat{\rho}_i(r,v,v_0,t)}
 {\rho_i(r,t)}
\end{equation}
on each lane $i$ as well as the {\em average interaction rate}
\begin{equation}
 \frac{1}{T_i^0} = \frac{1}{\rho_i(r,t)} 
 \int dv \, \tilde{\rho}_i(r,v,t) \!\!\int\limits_{w<v} \!\! dw \, (v - w)
 \tilde{\rho}_i(r,w,t)
\end{equation}
of a vehicle on lane $i$ with other vehicles on the same lane.

\subsection{Derivation of Moment Equations}

We are now ready for deriving the desired macroscopic traffic equations
from the gas-kinetic equation 
(\ref{Boltz1}) with (\ref{acc}), (\ref{adj}), (\ref{Boltz2}), (\ref{intrate}),
(\ref{spont}), and (\ref{w1}). Integration with
respect to $v_0$ gives us the {\em reduced gas-kinetic traffic equation}
\begin{mathletters}\label{red}\begin{eqnarray}
 \frac{\partial \tilde{\rho}_i}{\partial t}
 &+& \frac{\partial}{\partial r} (\tilde{\rho}_i v) + 
 \frac{\partial}{\partial v} \left( \tilde{\rho}_i 
 \frac{\tilde{V}_{0i}(v) - v}{\tau_i} \right) 
 \label{reda} \\
 &=& - (1-p_i) 
 \tilde{\rho}_i(r,v,t) \int dw \, (v - w) \tilde{\rho}_i(r,w,t)
 \label{redb} \\
 &+& p_{i-1}^+ \tilde{\rho}_{i-1}(r,v,t) \!\!\int\limits_{w < v} \!\! dw \, 
 (v - w) \tilde{\rho}_{i-1}(r,w,t) \label{redc} \\
 &+& p_{i+1}^- \tilde{\rho}_{i+1}(r,v,t) \!\!\int\limits_{w < v} \!\! dw \, 
 (v - w) \tilde{\rho}_{i+1}(r,w,t) \label{redd} \\
 &-& (p_i^+ + p_i^-) \tilde{\rho}_{i}(r,v,t) \!\!\int\limits_{w < v} \!\! dw
 \, (v - w) \tilde{\rho}_{i}(r,w,t) \label{rede} \\
 &+& \frac{1}{T_{i-1}^+} \tilde{\rho}_{i-1}(r,v,t)
 - \frac{1}{T_i^+} \tilde{\rho}_i(r,v,t) \label{redf} \\
 &+& \frac{1}{T_{i+1}^-} \tilde{\rho}_{i+1}(r,v,t)
 - \frac{1}{T_i^-} \tilde{\rho}_i(r,v,t) \label{redg} \\
 &+& \tilde{\nu}_i^+(r,v,t) - \tilde{\nu}_i^-(r,v,t) \label{redh}
\end{eqnarray}\end{mathletters}
with
\begin{equation}
 \tilde{V}_{0i}(v) \equiv \tilde{V}_{0i}(v;r,t)
 = \int dv_0 \, v_0 \frac{\hat{\rho}_i(r,v,v_0,t)}{\tilde{\rho}_i(r,v,t)}
\end{equation}
and
\begin{equation}
 \tilde{\nu}_i^\pm(r,v,t) = \int dv_0 \, \hat{\nu}_i^\pm(r,v,v_0,t) \, .
\end{equation}
In formula (\ref{red}), the deceleration term (\ref{redb}) 
stems from (\ref{intc}), 
the terms (\ref{redc}) to (\ref{rede}) reflecting immeditate overtaking come
from (\ref{inta}) and (\ref{intb}), and the lane-changing terms (\ref{redf}),
(\ref{redg}) originate from (\ref{w1}). 
The adaptation term $(\partial \hat{\rho}_i/\partial t)_{\rm ad}$
yields no contribution.
\par
We will now derive equations for the moments $\langle v^k \rangle$
by multiplying (\ref{red}) with $v^k$ and integrating with respect to $v$.
Due to
\begin{eqnarray}
 \int dv \, v^k \frac{\partial}{\partial v} \left( \tilde{\rho}_i 
 \frac{\tilde{V}_{0i}(v) - v}{\tau_i} \right) &=& - \int dv \, k v^{k-1}
 \left( \tilde{\rho}_i \frac{\tilde{V}_{0i}(v) - v}{\tau_i} \right)
\nonumber \\
 &=& - \frac{k \rho_i}{\tau_i} ( \langle v^{k-1} v_0 \rangle_i - \langle v^k
 \rangle_i ) 
\end{eqnarray}
and
\begin{equation}
 (1 - p_i) \int dv \, \tilde{\rho}_i(r,v,t) \int dw \, (wv^k - v^{k+1} )
 \tilde{\rho}_i(r,w,t) = (1-p_i) (\rho_i)^2 (\langle v \rangle_i
 \langle v^k \rangle_i - \langle v^{k+1} \rangle_i )
\end{equation}
we obtain the macroscopic moment equations
\begin{eqnarray}
 \frac{\partial}{\partial t} (\rho_i \langle v^k \rangle_i)
 + \frac{\partial}{\partial r} (\rho_i \langle v^{k+1} \rangle_i )
 &=& \frac{k \rho_i}{\tau_i} (\langle v^{k-1} v_0 \rangle_i - \langle v^k
 \rangle_i  ) \nonumber \\
 &+& (1-p_i) (\rho_i)^2 (\langle v \rangle_i
 \langle v^k \rangle_i - \langle v^{k+1} \rangle_i ) \nonumber \\
 &+& \frac{p_{i-1}^+}{T_{i-1}^0} \rho_{i-1} \langle v^k
 \rangle_{i-1} 
 - \frac{p_{i}^+}{T_{i}^0} \rho_{i} \langle v^k
 \rangle_{i}  \nonumber \\
 &+& \frac{p_{i+1}^-}{T_{i+1}^0} \rho_{i+1} \langle v^k
 \rangle_{i+1}  
 - \frac{p_{i}^-}{T_{i}^0} \rho_{i} \langle v^k \rangle_{i} 
 \nonumber \\
 &+& \frac{1}{T_{i-1}^+} \rho_{i-1} \langle v^k \rangle_{i-1} 
  - \frac{1}{T_{i}^+} \rho_{i} \langle v^k \rangle_{i} \nonumber \\
 &+& \frac{1}{T_{i+1}^-} \rho_{i+1} \langle v^k \rangle_{i+1} 
  - \frac{1}{T_{i}^-} \rho_{i} \langle v^k \rangle_{i} \nonumber \\
 &+& \nu_i^+(r,t) \langle v^k \rangle_i^+
  -  \nu_i^-(r,t) \langle v^k \rangle_i^- \, .
\label{mom1}
\end{eqnarray}
Here, we have introduced the notation 
\begin{equation}
 \langle v^k \rangle_i^\pm = \int dv \int dv_0 \, v^k 
 \frac{\hat{\nu}_i^\pm(r,v,v_0,t)} {\nu_i^\pm(r,t)}
= \int dv \, v^k \frac{\tilde{\nu}_i^\pm(r,v,t)}
 {\nu_i^\pm(r,t)} 
\end{equation}
and applied the approximation
\begin{eqnarray}
& & \hspace*{-2cm} 
 \int dv \, \tilde{\rho}_j(r,v,t)\!\! \int\limits_{w < v} \!\! dw \, v^k (v-w) 
 \tilde{\rho}_j(r,w,t) \nonumber \\
&\approx& \langle v^k \rangle_j 
 \int dv \, \tilde{\rho}_j(r,v,t)\!\! \int\limits_{w < v} \!\! dw \, (v-w) 
 \tilde{\rho}_j(r,w,t) = \langle v^k \rangle_j \frac{\rho_j(r,t)}{T_j^0} 
\end{eqnarray}
which is empirically justified due to the smallness of the
velocity distributions $P_j(v;r,t)$ (i.e. due to $\sqrt{\theta_j} \ll V_j$
\cite{Hel96,Hel97}).

\subsection{Fluid-Dynamic Multi-Lane Traffic Equations}

In order to derive dynamic equations for the densities
$\rho_i$ and average velocities $V_i$,
we need the relations
\begin{equation}
 \langle v^2 \rangle_i = \langle [ V_i + (v - V_i)]^2 \rangle_i
 = (V_i)^2 + 2 V_i \langle v - V_i \rangle_i + \langle (v-V_i)^2 \rangle_i
 = (V_i)^2 + \theta_i 
\end{equation}
and
\begin{equation}
 \rho_i \frac{\partial V_i}{\partial t} = \frac{\partial}{\partial t}
 ( \rho_i \langle v \rangle_i) - V_i \frac{\partial \rho_i}{\partial t} \, .
\end{equation}
Applying these and 
and using the abbreviations 
\begin{equation}
 \frac{1}{\tau_i^\pm} = \frac{p_i^\pm}{T_i^0} + \frac{1}{T_i^\pm} \, ,
\end{equation}
equation (\ref{mom1}) gives us the {\em density equations}
\begin{mathletters}\label{Dens}\begin{eqnarray}
 \frac{\partial \rho_i}{\partial t} + V_i \frac{\partial \rho_i}{\partial r}
 &=& - \rho_i \frac{\partial V_i}{\partial r} + \nu_i^+(r,t) - \nu_i^-(r,t)
 \label{Densa} \\
 &+& \frac{\rho_{i-1}}{\tau_{i-1}^+} - \frac{\rho_i}{\tau_i^+}
 + \frac{\rho_{i+1}}{\tau_{i+1}^-} - \frac{\rho_i}{\tau_i^-} \, .
\label{Densb}
\end{eqnarray}\end{mathletters}
This result is similar to previous multi-lane models.
However, by some lengthy but straightforward calculations we additionally
obtain the {\em velocity equations}
\begin{mathletters}\label{Veloc}\begin{eqnarray}
 \rho_i \frac{\partial V_i}{\partial t} + \rho_i V_i \frac{\partial V_i}
 {\partial r} &=& - \frac{\partial {\cal P}_i }{\partial r}
 + \frac{\rho_i}{\tau_i} (V_{i}^{\rm e} - V_i) 
\label{Veloca} \\
&+& \frac{\rho_{i-1}}{\tau_{i-1}^+} (V_{i-1} - V_i) 
 + \frac{\rho_{i+1}}{\tau_{i+1}^-} (V_{i+1} - V_i) \label{Velocb} \\
&+& \nu_i^+ (V_i^+ - V_i) - \nu_i^- (V_i^- - V_i ) \label{Velocc} 
 \end{eqnarray}\end{mathletters}
with the so-called {\em traffic pressures} \cite{PrHe71,Phi77,Phi79}
\begin{equation}
 {\cal P}_i = \rho_i \theta_i
\end{equation}
and the {\em equilibrium velocities}
\begin{equation}
 V_i^{\rm e} = V_{0i} - \tau_i (1-p_i) \rho_i \theta_i \, .
\end{equation}
Equation (\ref{Veloc}) corrects the phenomenological approach by
Michalopoulos et al. \cite{MiBe84}. 
The terms containing the rates $\nu_i^+$ and $\nu_i^-$ reflect entering and
leaving vehicles, respectively.
Whereas the terms (\ref{Densa}) and (\ref{Veloca}) correspond
to the effects of vehicle motion, of acceleration towards the drivers' 
desired velocities,
and of deceleration due to interactions, the terms (\ref{Densb}) and
(\ref{Velocb}) arise from overtaking and 
lane-changing maneuvers. (\ref{Velocb}) comes from differences between
the average velocities on neighboring lanes and tends to reduce them.
The term (\ref{Velocc}) has a similar
form and interpretation like (\ref{Velocb}). It
is only negligible if entering vehicles are able 
to adapt to the velocities on the merging lane and exiting vehicles initially
have an average velocity similar to that on the lane which they 
are leaving so that $V_i^\pm \approx V_i$.
\par
In order to close equations (\ref{Dens}) and (\ref{Veloc}),
we must specify the interaction rates $1/T_i^0$ and  
the variances $\theta_i$. Utilizing that the 
empirical velocity distributions $P_i(v;r,t)$ are approximately
{\em normally distributed} \cite{Hel96,MuPi71,Pam55,Phi77}, we have
\begin{equation}
 P_i(v;r,t) \approx {\textstyle\frac{1}{\sqrt{2\pi \theta_i(r,t)}}}
 \mbox{e}^{-[v - V_i(r,t)]^2/[2\theta_i(r,t)]} 
\end{equation} 
which implies 
\begin{equation}
 \frac{1}{T_i^0} \approx \rho_i \sqrt{\frac{\theta_i}{\pi}} \, .
\end{equation}
With a detailled theoretical and empirical analysis it can be
shown \cite{Hel95,Hel97} that the variances $\theta_i(r,t)$ can be well
approximated by equilibrium relations $\theta_i^{\rm e}(\rho_i)$
which are given by the implicit equation
\begin{equation}
 \theta_i^{\rm e}(\rho_i) \equiv \hat{\theta}_{i0} - 2\tau_i(\rho_i)
 [1-p_i^{\rm e}(\rho_i)] \frac{\rho_i [\theta_i^{\rm e}(\rho_i)]^{3/2}}{\sqrt{\pi}} \, .
\end{equation}
Here, $p_i^{\rm e}$ denotes the overtaking probability for vehicles on lane
$i$, when the densities $\rho_j$ on the different lanes $j$ are in equlibrium.
For the average desired velocities $V_{0i}$ we have
\begin{equation}
 V_{0i} \equiv V_{0i}(r,t) \approx \hat{V}_{0i}(r,\!\not t)
\end{equation}
since
\begin{equation}
 \hat{P}_{0i}(v_0;r,\!\not t) - P_{0i}(v_0;r,t) 
 \approx 0
\label{adiab}
\end{equation}
due to the smallness of $T_{\rm r}$. 

\section{Derivation and Simulation of a Reduced Multi-Lane Model}

The velocity equations are mainly needed to model the observed traffic
instabilities which lead to the spontaneous formation of 
stop-and-go waves at medium densities 
\cite{Kue84,KueRoe91,KeKo93,KeKo94,Hel96,KueBe93,Hel95c}. However, if one is
not interested in the density oscillations but only in the {\em average} 
temporal evolution of traffic flow, the velocity equations can be
eliminated. In order to do this, we will apply a method that has been
suggested by Sela and Goldhirsch \cite{SeGo}:
First, we introduce the {\em time averages}
\begin{equation}
 \overline{F}_i(r,t) = \frac{1}{\Delta T} 
 \!\!\!\!\!\int\limits_{t-\Delta T/2}^{t+\Delta T/2}\!\!\!\!\!
 dt \, F_i(r,t)
\end{equation}
over the least common multiple $\Delta T$ of the occuring 
oscillation periods $\Delta T_i$. Then, the quantities
$\overline{\rho}(r,t)$ and $\overline{V}(r,t)$ will describe the 
{\em coarse-grained} traffic dynamics, in other words: the traffic dynamics
on a slow time scale. Additionally, the time averages of the
total time derivatives $d\rho_i/dt$ ad $dV_i/dt$ will approximately vanish:
\begin{equation}
 \overline{\frac{d\rho_i}{dt}} \equiv \overline{\frac{\partial \rho_i}
 {\partial t}} + \overline{V_i \frac{\partial \rho_i}{\partial r}}
 \approx 0 \, , \qquad 
 \overline{\frac{dV_i}{dt}} \equiv \overline{\frac{\partial V_i}
 {\partial t}} + \overline{V_i \frac{\partial V_i}{\partial r}}
 \approx 0 \, .
\label{this}
\end{equation}
This corresponds to the assumption that, in coordinate systems moving 
with velocities $V_i(r,t)$, the densities $\rho_i(r,t)$ and
velocities $V_i(r,t)$ oscillate around their (slowly changing) equilibrium  
values.
\par
Now, we approximate time averages $\overline{F_i(\rho_i,V_i)}$
of density- and velocity-dependent functions $F_i(\rho_i,V_i)$
by a series in spatial derivatives of $\overline{\rho_i}$ and
$\overline{V_i}$. For our purposes it is sufficient to truncate
the expansion after the first order \cite{Hel97}:
\begin{equation}
 \overline{F_i(\rho_i,V_i)} \approx F_{00}(\overline{\rho_i},\overline{V_i})
 + F_{10}(\overline{\rho_i},\overline{V_i}) \frac{\partial \overline{\rho_i}}
 {\partial r}
 + F_{01}(\overline{\rho_i},\overline{V_i}) \frac{\partial \overline{V_i}}
 {\partial r} \, .
\label{expans}
\end{equation}
With this and (\ref{this}) we obtain from the time average of 
velocity equations (\ref{Veloc}):
\begin{equation}
 \overline{V_i} = \frac{\displaystyle \frac{\overline{\rho_i}}
 {\tau_i} V_i^{\rm e}(\overline{\rho_i}) 
 + \frac{\overline{\rho_{i-1}}}{\tau_{i-1}^+} \overline{V_{i-1}}
 + \frac{\overline{\rho_{i+1}}}{\tau_{i+1}^-} \overline{V_{i+1}} 
 - \frac{\partial {\cal P}_i(\overline{\rho_i})}
 {\partial \overline{\rho_i}} \frac{\partial \overline{\rho_i}}{\partial r}}
 {\displaystyle \frac{\overline{\rho_i}}{\tau_i}
 + \frac{\overline{\rho_{i-1}}}{\tau_{i-1}^+}
 + \frac{\overline{\rho_{i+1}}}{\tau_{i+1}^-} }
\, .
\label{solving}
\end{equation}
Here, we have restricted our considerations to the case of a freeway without entrances and
exits. Resolving (\ref{solving}) with respect to $\overline{V_i}$ leads
to a relation of the form
\begin{equation}
 \overline{V_i} 
 = {\cal V}_i(\{\overline{\rho_j} \}) -
 \sum_k \frac{{\cal D}_k(\{\overline{\rho_j}\})}{\overline{\rho_i}}
 \frac{\partial \overline{\rho_k}}{\partial r}  
\end{equation}
which only depends on the densities $\overline{\rho_j}$ and their
gradients. Inserting this into the density equations (\ref{Dens})
finally leads to the reduced equations
\begin{equation}
 \frac{\partial \overline{\rho_i}}{\partial t}
 + \frac{\partial}{\partial r} 
 [ \overline{\rho_i} {\cal V}_i(\{ \overline{\rho_j} \} ) ]
 = \sum_k \frac{\partial}{\partial r} \left[
 {\cal D}_k(\{\overline{\rho_j}\}) \frac{\partial \overline{\rho_k}}{\partial
     r} \right] + \frac{\overline{\rho_{i-1}}}{\tau_{i-1}^+} 
 - \frac{\overline{\rho_i}}{\tau_i^+}
 + \frac{\overline{\rho_{i+1}}}{\tau_{i+1}^-} 
 - \frac{\overline{\rho_i}}{\tau_i^-} \, .
\end{equation}
If we neglect products of spatial derivatives we end up with
the coupled {\em Burgers equations} \cite{Whi74}
\begin{equation}
 \frac{\partial \overline{\rho_i}}{\partial t}
 + \frac{\partial}{\partial r} 
 [ \overline{\rho_i} {\cal V}_i(\{ \overline{\rho_j} \} ) ]
 = \sum_k 
 {\cal D}_k(\{\overline{\rho_j}\}) \frac{\partial^2 \overline{\rho_k}}{\partial
     r^2} + \frac{\overline{\rho_{i-1}}}{\tau_{i-1}^+} 
 - \frac{\overline{\rho_i}}{\tau_i^+}
 + \frac{\overline{\rho_{i+1}}}{\tau_{i+1}^-} 
 - \frac{\overline{\rho_i}}{\tau_i^-} \, .
\label{Burg}
\end{equation}
On the rigth-hand side of this equation we have a sum of {\em diffusion terms}
with density-dependent diffusion functions ${\cal D}_k$. These cause
a smoothing of sudden density changes and prevent the formation of shock
waves. This is the reason why density gradients and especially
products of spatial derivatives are normally negligible which justifies
the approximation made with equation 
(\ref{Burg}).
Apart from this, the diffusion terms are very helpful for efficient
and stable numerical integration schemes.
\par
\begin{figure}[htbp]
\unitlength10mm
\begin{center}
\begin{picture}(16,10.6)(0,-0.8)
\put(0,9.8){\epsfig{height=16\unitlength, width=9.8\unitlength, angle=-90, 
      bbllx=50pt, bblly=50pt, bburx=554pt, bbury=770pt, 
      file=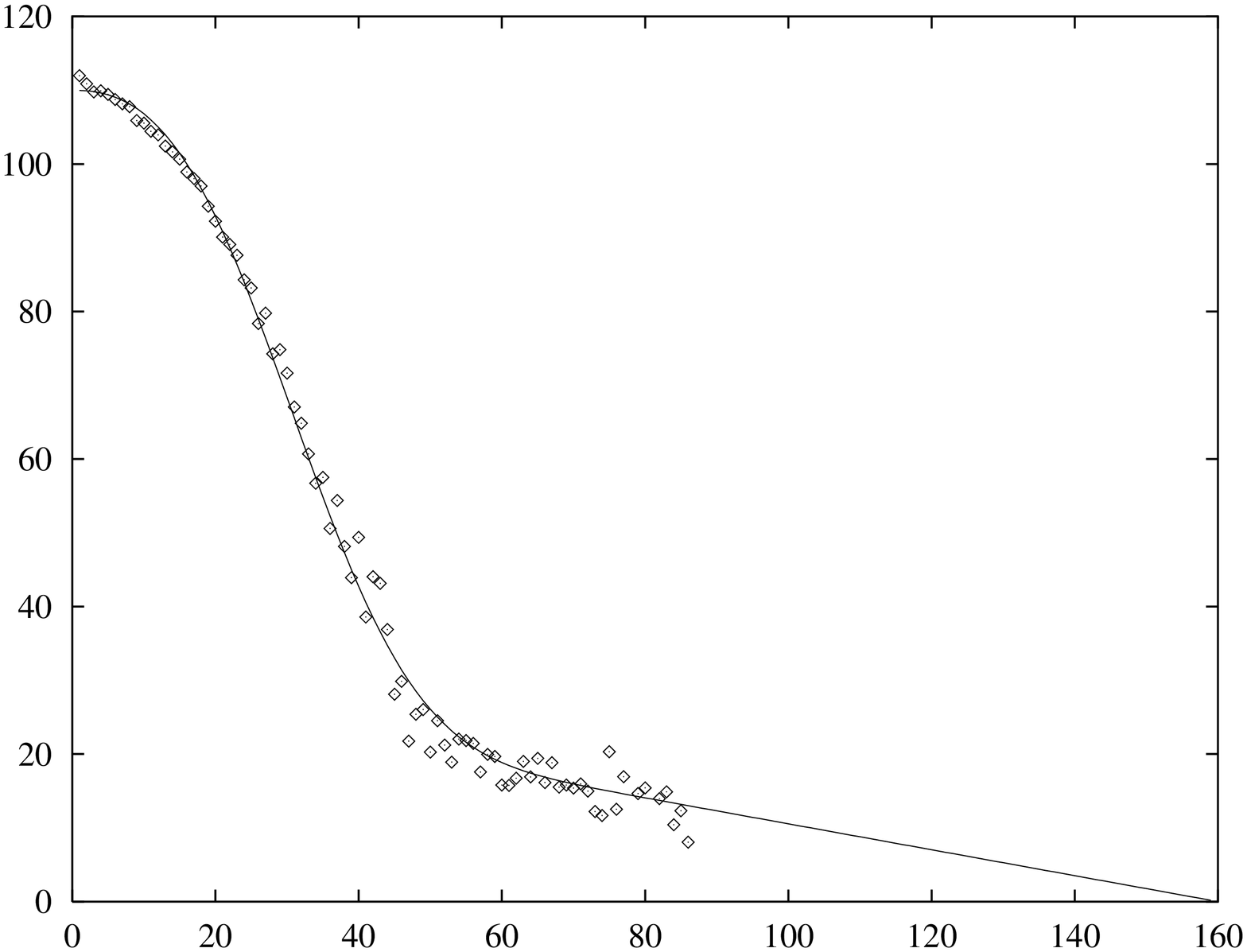}}
\put(8.7,-0.6){\makebox(0,0){\footnotesize $\rho_i$ (vehicles/km lane)}}
\put(0.2,5.1){\makebox(0,0){\rotate[l]{\hbox{\footnotesize $V_i^{\rm e}
(\rho_i)$ (km/h)}}}}
\end{picture}
\end{center}
\caption[]{The chosen velocity-density relation $V_i^{\rm e}(\rho_i)$ (---)
and the corresponding empirical data from the Dutch autobahn A9
($\Diamond$).} 
\label{empvel}
\end{figure}
\begin{figure}[h]
\unitlength10mm
\begin{center}
\begin{picture}(16,10.6)(0,-0.8)
\put(0,9.8){\epsfig{height=16\unitlength, width=9.8\unitlength, angle=-90, 
      bbllx=50pt, bblly=50pt, bburx=554pt, bbury=770pt, 
      file=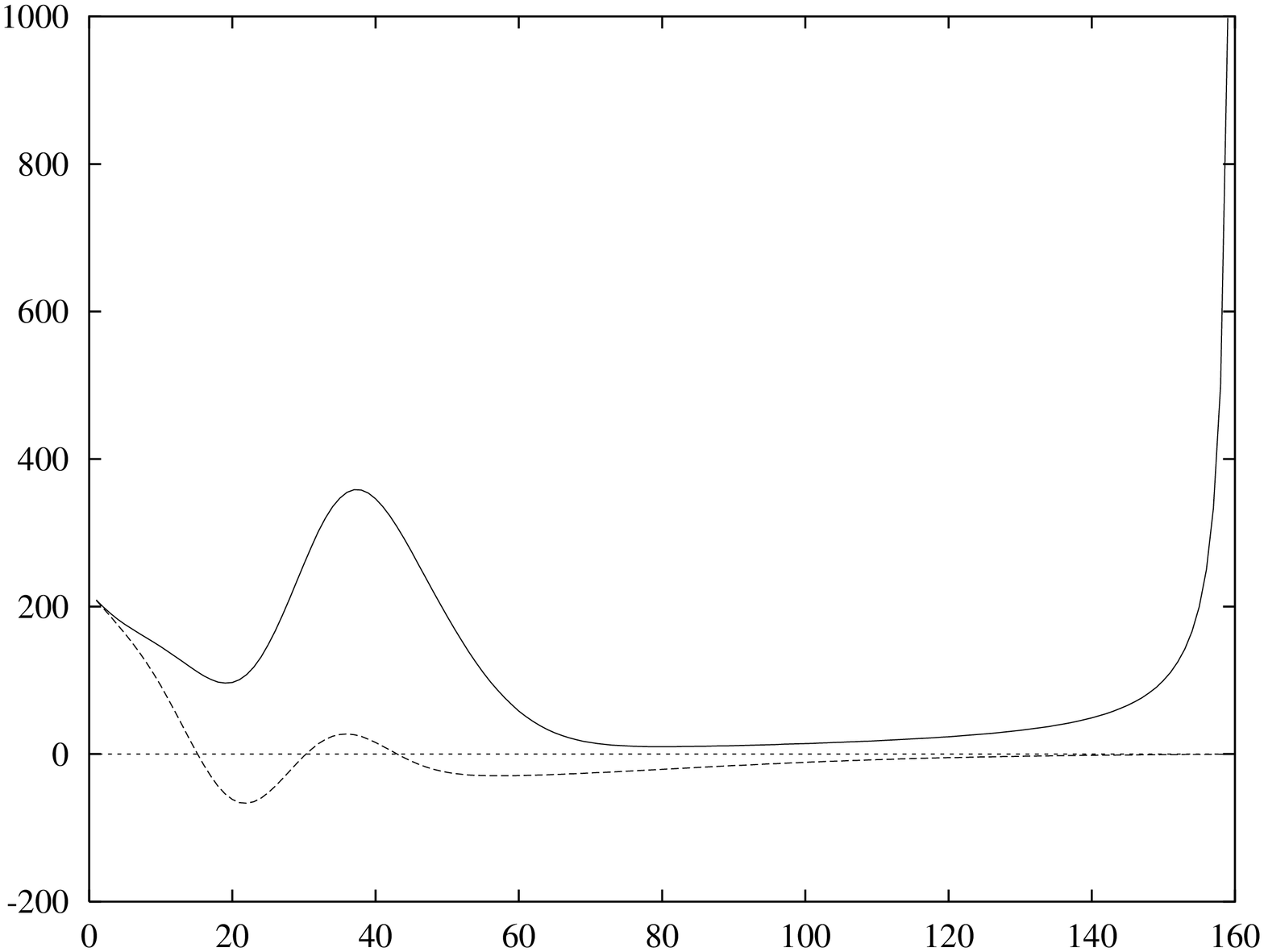}}
\put(8.7,-0.6){\makebox(0,0){\footnotesize $\rho_i$ (vehicles/km lane)}}
\put(0,5.1){\makebox(0,0){\rotate[l]{\hbox{\footnotesize $\displaystyle
\frac{\partial {\cal P}_i}{\partial 
\rho_i}$ (km$^2$/h$^2$)}}}}
\end{picture}
\end{center}
\caption[]{Representation of the density-gradients of the
{\em idealized} traffic pressure ${\cal P}_i = \rho \theta_i^{\rm e}(\rho)$ 
of point-like vehicles (--~--) and the {\em corrected} pressure relation 
(---) which takes into account their finite space requirements.
Obviously, the increase of the corrected traffic pressure
with density and the corrected traffic pressure itself diverges at
the maximum density $\rho_{\rm max}$, so that the latter cannot be exceeded.
For this reason, the diffusion funktions ${\cal D}_k$ also diverge
for $\rho_k \rightarrow \rho_{\rm max}$.
The pressure relations have been reconstructed from empirical
data by means of theoretical relations \protect\cite{Hel96,Hel97}.}
\label{dP}
\end{figure}

We will now focus on the simulation of multi-lane traffic. As an example,
we investigate a two-lane autobahn (i.e. $i \in \{1,2\}$). For reasons of 
simplicity, on both lanes the velocity-density relations $V_i^{\rm e}(\rho_i)$
and pressure relations ${\cal P}_i(\rho_i)$ will be chosen identically.
This is at least justified for congested traffic (with a density of
30 vehicles per kilometer and lane or more). The corresponding 
relations are depicted in Figures~\ref{empvel} and \ref{dP}. 
They have been constructed
from empirical data of the Dutch highway A9 between Haarlem and Amsterdam
with a speed limit of 120\,km/h and take into account corrections of
the traffic equations for high densities (for details cf. 
Refs.~\cite{Hel96,Hel97}). 
\par
The lane-changing rates 
$1/\tau_i^\pm$ are chosen in accordance
with an empirically validated model \cite{Spar}:
\begin{equation}
 \frac{1}{\tau_i^\pm} = \beta_i^\pm \overline{\rho_i}
 ( \rho_{\rm max} - \overline{\rho_{i\pm 1}} ) \, .
\end{equation}
Therefore, $1/\tau_i^\pm$ is proportional to the vehicle density 
$\overline{\rho_i}$ which reflects the grade of obstruction by slower 
vehicles on lane $i$. The factor
$(\rho_{\rm max} - \overline{\rho_{i\pm 1}})$ reflects that vehicles
can change to the neighboring lane $i\pm 1$ less frequently the more
the density on it reaches the {\em maximum density} $\rho_{\rm max}$.
For German autobahns the parameters $\beta_i^\pm$ have the following values:
\begin{equation}
 \beta_1^+ = 0.176 \cdot 10^{-3} \, , \qquad
 \beta_2^- = 0.056 \cdot 10^{-3} \, , \qquad 
 \beta_1^- = \beta_2^+ = 0 \, .
\end{equation}
The relation $\beta_1^+ > \beta_2^-$ originates from the fact that
the left lane is preferred in Germany, since overtaking is forbidden on
the right-hand lane.
\par
\begin{figure}[htbp]
\unitlength1.4cm
\begin{center}
\begin{picture}(10,7)(0,0)
\put(0,6){\epsfig{height=10\unitlength, angle=-90, 
      bbllx=6cm, bblly=5cm, bburx=18cm, bbury=24cm, 
      file=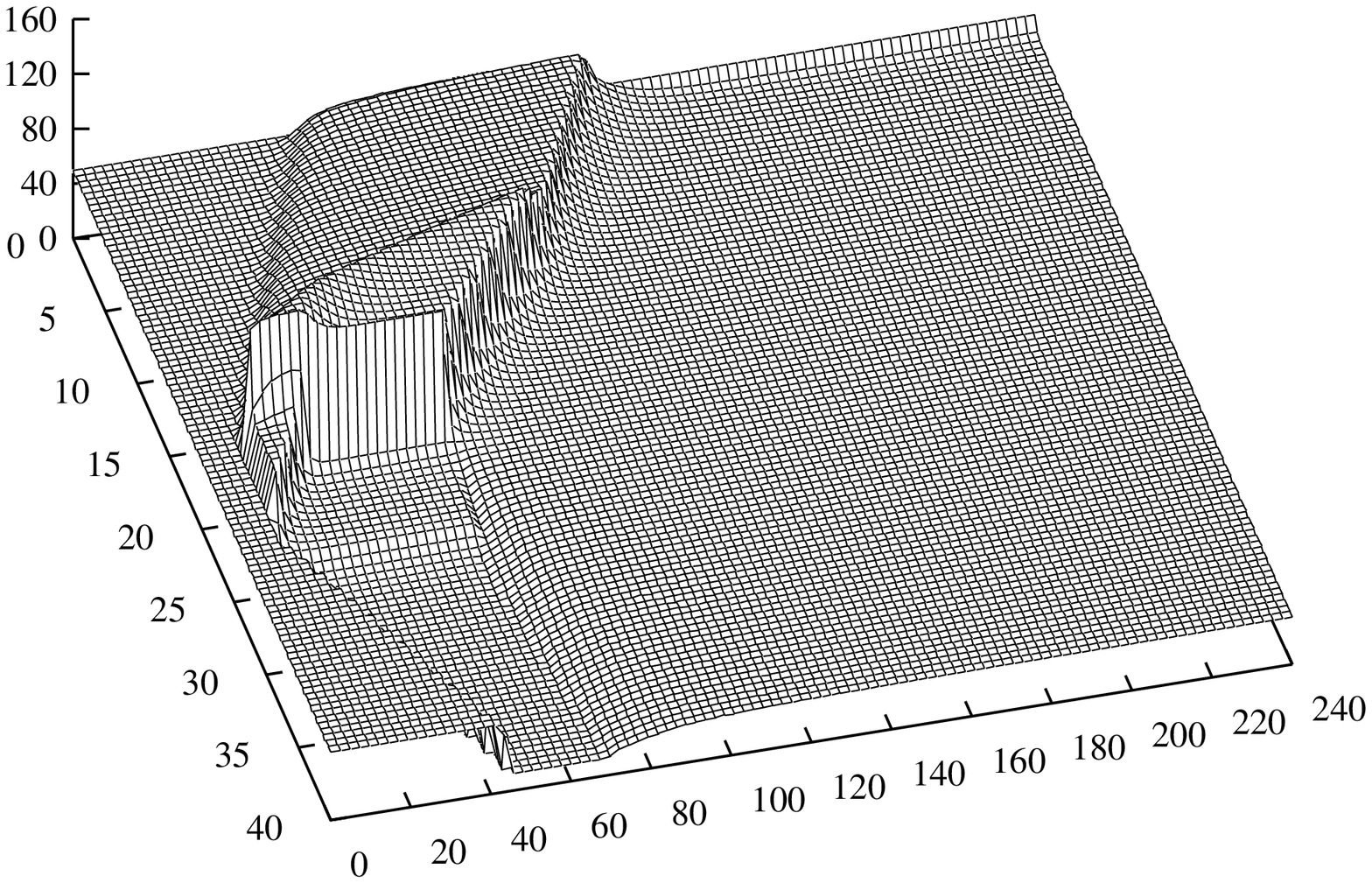}}
\put(0.65,6.3){\makebox(0,0){$\overline{\rho_2}(r,t)$ (vehicles/km lane)}}
\put(6.3,-0.05){\makebox(0,0){$t$ (min)}}
\put(0.1,2.15){\makebox(0,0){$r$ (km)}}
\end{picture}
\vskip1em
\begin{picture}(10,7)(0,0)
\put(0,6){\epsfig{height=10\unitlength, angle=-90, 
      bbllx=6cm, bblly=5cm, bburx=18cm, bbury=24cm, 
      file=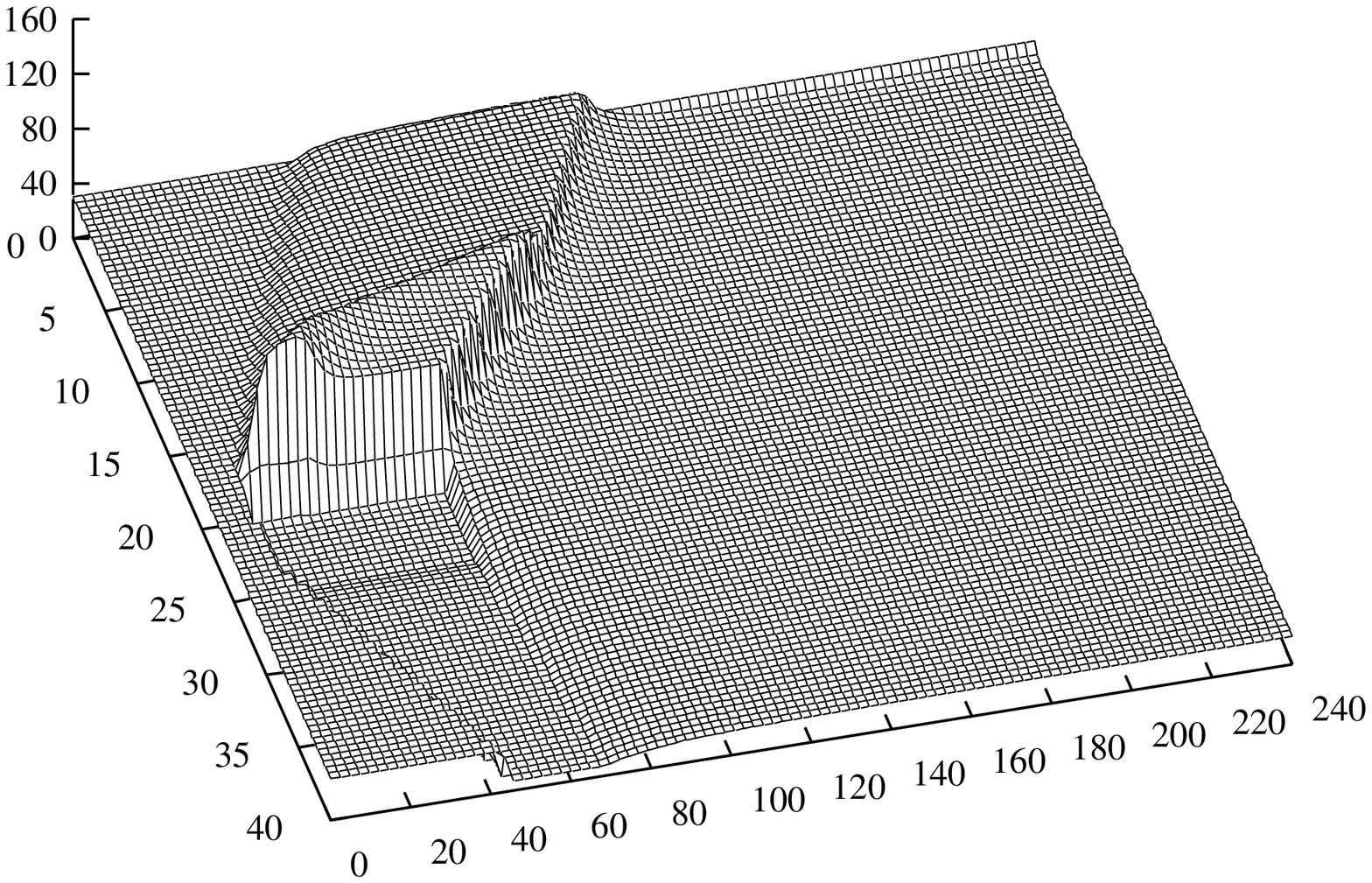}}
\put(0.65,6.3){\makebox(0,0){$\overline{\rho_1}(r,t)$ (vehicles/km lane)}}
\put(6.3,-0.05){\makebox(0,0){$t$ (min)}}
\put(0.1,2.15){\makebox(0,0){$r$ (km)}}
\end{picture}
\vskip1em
\end{center}
\caption[]{Spatio-temporal evolution of the time-averaged densities
on a two-lane freeway stretch of 40\,km length with open boundary conditions
in the case of an overloaded temporary bottleneck situation (above: left lane;
below: right lane). The right lane is closed between 
$t=10$\,min und $t=60$\,min on the stretch between
$r_0 = 21$\,km and $r_1=25$\,km, and the vehicles on lane 1 try to get on 
lane 2 beginning at $r_0-\Delta r = 20$\,km. This suddenly increases the
density on the left lane in the region of the bottleneck, whereas the right
lane becomes empty. Already after a
short time the extremest clustering develops at the beginning of the 
bottleneck, where the vehicles of the closed lane try to squeeze in
the left lane. Since the capacity of the remaining lane is smaller than
the total traffic volume, the left lane becomes overloaded.
For this reason a congestion running upstream (tailback) 
builds up on {\it both} lanes.
On the other hand, the density after the bottleneck, where two lanes are
available again, is smaller than in front of it so that the vehicles can
accelerate there. As a consequence, the traffic situation already recovers in
the course of the bottleneck. At $t=60$\,min, the lane closure is lifted
and the traffic jam disappears.
(Note: Due to the different lane-changing rates the equilibrium density is
somewhat greater than 40 vehicles per kilometer and lane on the
left lane and somewhat smaller on the right lane.)}
\label{and2}
\end{figure}
\begin{figure}[htbp]
\unitlength1.4cm
\begin{center}
\begin{picture}(10,7)(0,0)
\put(0,6){\epsfig{height=10\unitlength, angle=-90, 
      bbllx=6cm, bblly=5cm, bburx=18cm, bbury=24cm, 
      file=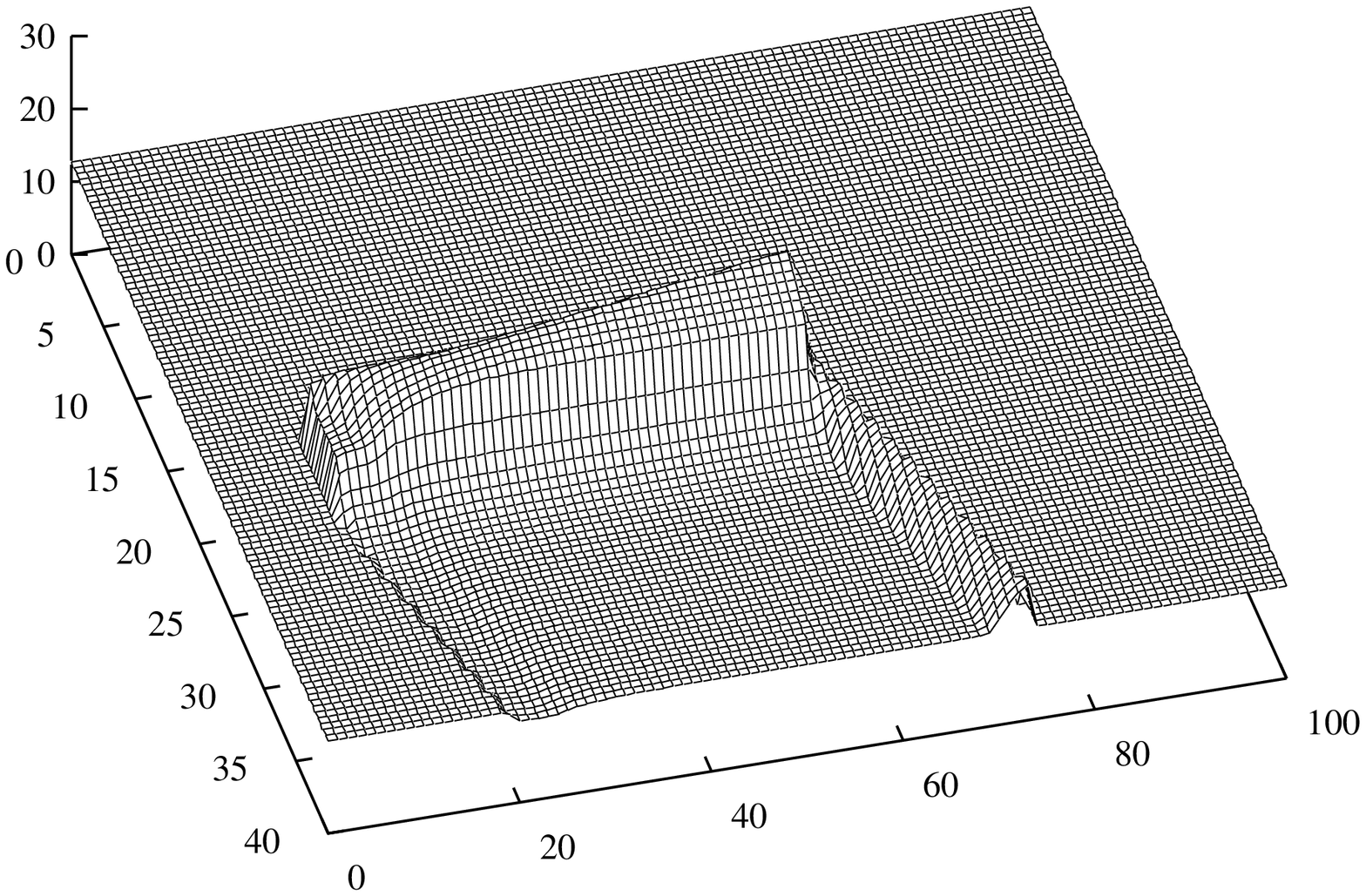}}
\put(0.65,6.3){\makebox(0,0){$\overline{\rho_2}(r,t)$ (vehicles/km lane)}}
\put(6.3,-0.05){\makebox(0,0){$t$ (min)}}
\put(0.1,2.15){\makebox(0,0){$r$ (km)}}
\end{picture}
\vskip1em
\begin{picture}(10,7)(0,0)
\put(0,6){\epsfig{height=10\unitlength, angle=-90, 
      bbllx=6cm, bblly=5cm, bburx=18cm, bbury=24cm, 
      file=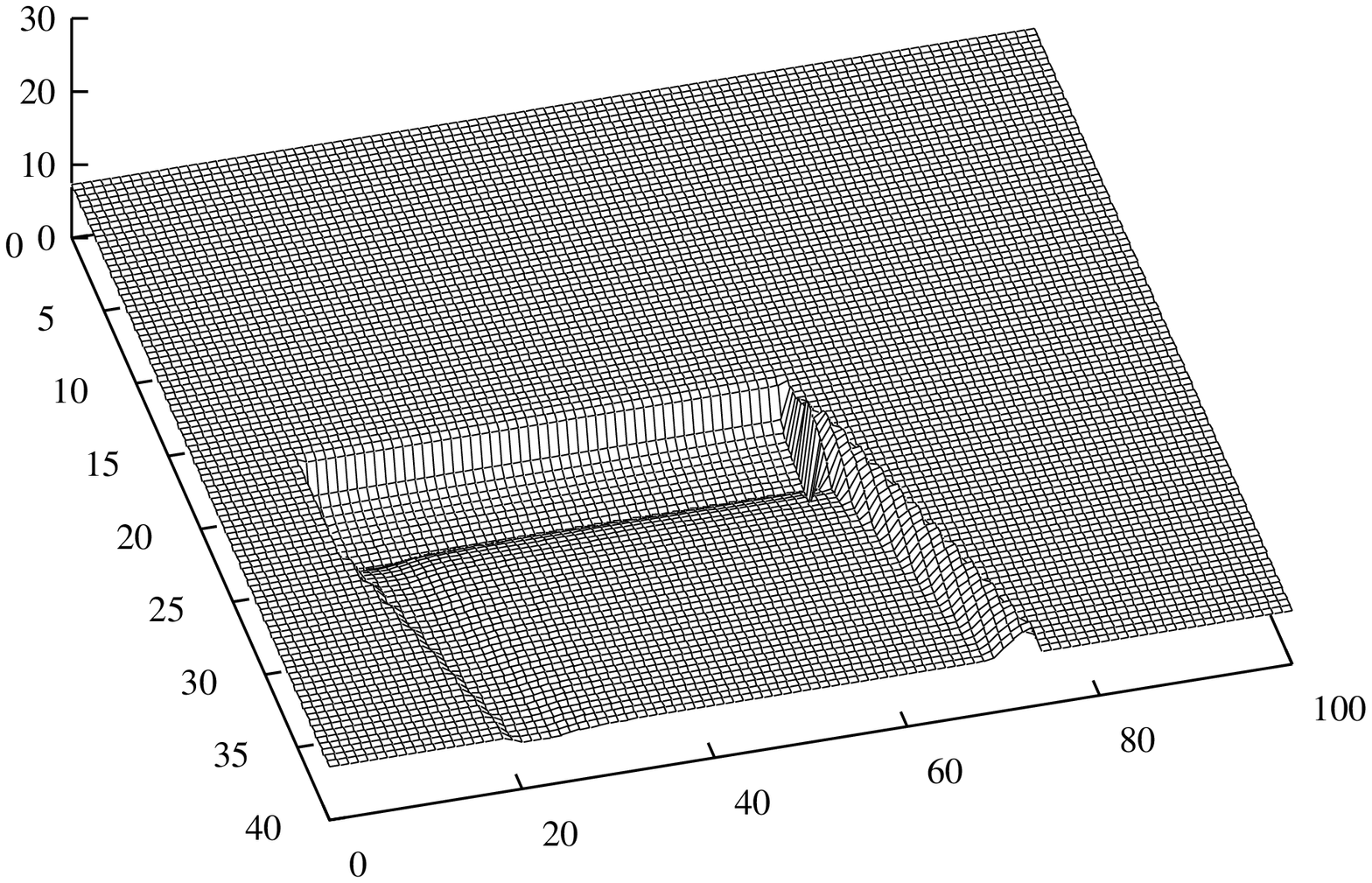}}
\put(0.65,6.3){\makebox(0,0){$\overline{\rho}_1(r,t)$ (vehicles/km lane)}}  
\put(6.3,-0.05){\makebox(0,0){$t$ (min)}}
\put(0.1,2.15){\makebox(0,0){$r$ (km)}}
\end{picture}
\vskip1em
\end{center}
\caption[]{Same as Figure \protect\ref{and2}, but for a bottleneck
situation below capacity. At an average density of 10 vehicles per
kilometer and lane the traffic capacity of one lane is large enough to cope
with the total traffic volume. Therefore, no congestion running upstream
builds up, but on the left lane the density is increased 
in the region of the bottleneck. This clustering only slowly dissolves in
the course of the road. After the lane closure is lifted, 
the traffic jam, which was previously localized at the bottleneck,
causes a damped density wave. This propagates 
along the freeway with a velocity that is slower
than the average vehicle velocity. Due to lane-changing maneuvers, 
the right lane also develops a propagating density wave.}
\label{and3}
\end{figure}

Bottleneck situations can be simulated in the following way: We will assume 
that the right lane is closed between places $r_0$ and $r_1$. Then, 
the lane-changing rate $\beta_1^+$ will be considerably increased, but
$\beta_2^-$ will be zero on this stretch and already a certain interval
$\Delta r$ before (i.e. for $r_0 - \Delta r \le r \le r_1$). 
$\beta_1^+$ and $\Delta r$ must be chosen sufficiently
large so that the right lane is empty up to the beginning $r_0$ of 
the bottleneck. Simulation results for the traffic dynamics above and
below capacity are presented in Figures \ref{and2}
and \ref{and3}, respectively.

\section{Summary and Outlook}

In this paper we have derived a macroscopic traffic model for uni-directional
multi-lane roads. Our considerations started from plausible assumptions about
the behavior of driver-vehicle units regarding acceleration, overtaking,
deceleration, and lane-changing maneuvers. The 
resulting gas-kinetic traffic model is a generalization of Paveri-Fontana's
Boltzmann-like traffic equation. It can be extended to situations where
different vehicle types or driving styles are to be investigated \cite{Hel97}.
\par
The gas-kinetic traffic equations not only allow to derive dynamic equations for
the vehicle density on each lane, but also for the
average velocity. In this way we were able to extend and correct previous
phenomenological multi-lane models. Overtaking and lane-changing maneuvers
are explicitly taken into account, so that the interactions between
neighboring lanes are included.
\par
We have then eliminated the velocity equations in order to obtain a reduced
model that allows efficient computer simulations. The resulting density
equations describe the 
average temporal evolution of traffic on a slow time
scale. They contain diffusion terms which diverge at maximum density 
$\rho_{\rm max}$ if
the finite space requirements of vehicles are taken into account. This
guarantees that $\rho_{\rm max}$ cannot be exceeded and density shocks
are smoothed out. The latter is important for realistic results and
stable numerical integration schemes. Finally, the reduced multi-lane
traffic model has been applied to the difficult case of bottleneck situations.
The computational results were very plausible. Consequently, the model can be
used to investigate a number of questions concerning the optimization of 
traffic flow:
\begin{enumerate}
\item In which way does on-ramp traffic influence and destabilize the traffic
flow on the other lanes? How does the destabilization effect depend on the
traffic volume, the length of the on-ramp lane, the total lane number, etc.?
\item In case of a reduction of the number of lanes, is it better to close the
left-most or the right-most lane? 
\item Is the organization of American freeways or of European autobahns more
efficient, or is it a suitable mixture of both? Remember that American
freeways are characterized by uniform speed limits and the fact that
overtaking as well as lane changing is allowed on both neighboring lanes.
In contrast, on European autobahns 
often no speed limit is prescribed (at least in
Germany) and average velocity normally increases with growing lane number
since overtaking is only allowed on the left-hand lane.
\item In which traffic situations do stay-in-lane recommendations increase the
efficiency of roads?
\end{enumerate}

\subsection*{Acknowledgment}

The authors want to thank Henk Taale and the Dutch Ministry of Transport,
Public Works and Water Management for supplying the empirical traffic
data.

\clearpage
\clearpage
\clearpage
\end{document}